\documentclass[12pt]{article}
\usepackage{epsfig}
\begin{document}
\begin{center}
{
 \renewcommand\thefootnote{$\dagger$}

{\bf\large OVERVIEW OF THE COMPETE PROGRAM\footnote{presented at 
the Second International "Cetraro" Workshop \& NATO
Advanced Research Workshop "Diffraction 2002", Alushta, Crimea, 
Ukraine, August 31 - September 6, 2002.}}\\
{}~\\
}
\setcounter{footnote}{0}

V.~V.~Ezhela\footnote{COMPAS group, IHEP, Protvino, Russia},
J.~R.~Cudell\footnote{Institut de Physique,
B\^at. B5, Universit\'e de Li\`ege, Sart Tilman, B4000 Li\`ege, Belgium},
P.~Gauron\footnote{LPNHE(Unit\'e de Recherche des Universit\'es
Paris 6 et Paris 7, Associ\'ee au CNRS)-Theory Group, Universit\'e
Pierre et Marie Curie, Tour 12 E3, 4 Place Jussieu, 75252
Paris Cedex 05, France}, 
K.~Kang\footnote{Physics Department, Brown University, Providence, RI, 
U.S.A.},  
S.~K.~Kang\footnote{ Department of Physics, Seoul 
National University, Seoul 131-717, Korea},
Yu.~V.~Kuyanov$^1$,
A.~Lengyel\footnote{Institute of Electron Physics, Universitetska  
21, UA-88000 Uzhgorod, Ukraine},
K.~S.~Lugovsky$^1$, 
S.~B.~Lugovsky$^1$,
V.~S.~Lugovsky$^1$,
E.~Martynov$^{2,}$\footnote{On leave from Bogolyubov Institute for 
Theoretical Physics, 03143 Kiev, Ukraine},
B.~Nicolescu$^3$,   
E.~A.~Razuvaev$^1$, M.~Yu.~Sapunov$^1$, 
O.~Selyugin$^{2,}$\footnote{ On leave from Bogoliubov Theoretical
Laboratory, JINR, 141980 Dubna, Moscow Region, Russia}, 
N.~P.~Tkachenko$^1$,
M.~R.~Whalley\footnote{HEPDATA group, Durham University, 
Durham, United Kingdom}, 
O.~V.~Zenin$^1$ \\
{}~\\

 \renewcommand\thefootnote{$\diamond$}
{COMPETE\footnote{COmputerised Models, Parameter Evaluation for Theory and
Experiment.
} collaboration}
\end{center}

\section*{Introduction}
Nowadays, scientific databases have become the bread-and-butter 
of particle physicists.
They are used not only for citation and publication
\cite{SPIRES,arxiv,prola,rfbr}, but also for
access to data compilations \cite{PDG,COMPAS,HEPDATA} and
for the determination of the best parameters of currently accepted models
\cite{PDG,COMPAS,CODATA}. These databases provide
inestimable tools
as they organize our knowledge in a coherent and
trustworthy picture.
They have lead not only to published works such as
the Review of Particle Physics \cite{PDG2002}, but also to web interfaces,
and reference data
compilations available in a computerized format readily
usable by physicists. It should be pointed out at this point
that one is far from using the full power of the web, as cross-linking
between various databases and interactive interfaces are only
sketchy. Part of the problem comes from the absence of a common 
repository or environment.

One must also stress that this crucial activity is not a given, and that
these databases must be maintained and checked repeatedly to insure
the accuracy of their content. In fact, we run the risk to loose some
or all of the information contained in them, as the maintainers are
getting older. There is a need for teaching the systematization and
evaluation of data in the standard physics curriculum, need which so
far has totally been overlooked. We stress the importance of summer
schools and workshops dedicated to data systematization.

The COMPETE collaboration aims at motivating data maintenance via the
interfacing of theory and experiment at the database level,
thus providing a complete picture
of the phenomenology describing a given subclass of phenomena. The
database concept then needs to be supplemented by 
a ``model-base" \cite{CPC1984}. Such
an object enables one not only to decide what the best description may be, but
also to discern what potential problems exist in the data. The
systematization of such a cross-fertilization between models and
data, which is at
the core of physics, results in what we shall call an
``object of knowledge", containing both factual and theoretical
information, and presumably becoming the point at which all existing
information resources on a given problem could converge.

There are many advantages to such a global approach.
First of all, the maintenance of a data set is not a static task: it
needs to be motivated by physics. Discrepancies between
models and data call for checks, and often those checks lead to a
new data set, where published errata in data are fixed and preliminary 
data are removed. A clear
example of such improvements can be found in the total cross section
data set. Furthermore, at times such studies show that there may be
problems in the experimental analysis itself. For instance, in the
analysis of the $\rho$ parameter, the systematic error resulting from
the use of a specific model is usually neglected. A general re-analysis
of these data is therefore needed, or a different treatment of systematic
errors may be brought in.

The second advantage is obviously that one can have a common testing
ground for theories and models, so that all the details of the comparison
are under control. This means that it becomes possible, for a given
set of assumptions, to define the best models reproducing a given
set of data. In this respect, as many models have to be tested,
and as the usual ``best fit" criterion, i.e. lowest $\chi^2/dof$, is
not fully satisfactory, we have developed a set of procedures that
enable artificial intelligence decisions, simulating to some extent
a physicist's intuition and taste.

Thirdly, it is obvious that an extensive theoretical database can
be used to plan new experiments, and to predict various quantities.
The automated treatment of a large number of models and theories
enables us to quote a theoretical error, which gives the interval
in which existing models can reproduce experimental results, and
to determine the sensitivity needed to discriminate between various models.

Finally, as new data come in, one can very quickly decide on their
theoretical impact, and hence immediately evaluate the need
for new physics ideas.

As we want to treat a large amount of data and many models, computer
technology constitutes an important part of our activity. We have concentrated
on the elaboration of artificial intelligence decision-making algorithms, as
well as on the delivery of computer tools for the end-user: these
include web summaries of results, web calculators of various quantities
for the best models, and of course computer-readable data-sets and Fortran
codes. Finally, the consideration of several different physics  problems
brings in the need to interface various objects of knowledge. The
interconnection and compatibility of these is an important constraint.
Further linkage with existing
databases, such as PDG \cite{PDG}, COMPAS \cite{COMPAS}, and HEPDATA
\cite{HEPDATA} is being developed or planned.

\subsection*{Methodology}
Our work is based on the following information model: theoretical descriptions
and data
are arranged in bases, which are then interfaced. This cross assessment leads
to a ranking of models, and to an evaluation of data: some
models globally reproduce data and some don't, some
data seem incompatible with all models or with similar data. At this
point, a selection of acceptable models and of acceptable data is
made. The models are then ranked according not only to their
$\chi^2/dof$, but also to their number of parameters, stability,
extendability to other data, etc. The best models are kept,
and organized into an object of knowledge. One can then make predictions
based on these few best models, and evaluate theoretical errors from
all acceptable models. The data set can also be re-evaluated,
and after a new data set is produced, one can re-iterate the above procedure.
The next step is then to find models that can accommodate more data
and once such new models are proposed, one can iterate again.

It is worth pointing out here the problems directly linked with data and
parameters.
First of all, contradictory data lead to sizable uncertainties. One
way to handle these is to use a Birge factor, renormalizing the $\chi^2/dof$ to
1. Another way is to re-normalize a given data set, and assign to
this operation a penalty factor. Finally, it is also possible to
shift data sets within their systematic errors in order to obtain
the best data set overall. Each method has its problems and advantages,
and no overall best method has been found so far.
Secondly, one must stress that, besides the usual statistical and
systematic errors, one should independently quote a theoretical error
which may not be combined with the experimental systematics. Finally,
it is important for a given set of parameters to indicate their area
of applicability, {\it e.g.} often high-$Q^2$ or high-$s$ models are used
outside their area of validity.

The organization of work within the collaboration
is similar to that of the Particle Data Group:
each member of the collaboration has access to the current object of knowledge
which gets released to the community once a year, in the form of
computer-usable files, and web-accessible notes.
Part of the collaboration
is devoted to the finding of new data and models in the literature, as well as
to their encoding. Other people check the accuracy of the encoding.
The study and elaboration of the object of knowledge is done under
the guidance of a few developers, whose work then gets partially or fully
verified by the rest of the team in charge of that study. The web interface and
tools are then developed or updated by another part of the collaboration.

The organization of the object of knowledge itself goes as follows: a
compilation of data is interfaced with a compilation of models, typically
kept as a set of Mathematica routines (which can then be used to produce
Fortran or C). The conclusions of the cross-assessment are then fed into
a program devoted to predictions, and freely executable via the web.
Tables of predictions then become available. Another module gathers
the information obtained from the cross-assessment and makes it available to
the collaboration for cross-checks. Finally, another module uses
the citation databases to track new work that refers to our existing databases.
\newpage
\section*{Results}
The results we have obtained so far fall within two main categories:
the first concerns the tools that we have developed, which could be
used by others in a wide variety of tasks, the second concerns the physics
conclusions which we have reached.

\subsection*{Tools}
\subsubsection*{Elements of the artificial intelligence}
The usual indicator $\chi^2/dof$ is certainly an important measure
of the quality of a fit. However, it does not give us all the relevant
information to choose the best models. We have developed
\cite{PRD2002,JRC}, in the
context of fits to soft data, a series of other indicators that enable
us to study numerically some of the aspects of fits which so far had only
received a qualitative treatment. Models usually rely on some approximation
which breaks down in some region. For instance, in DIS, the starting
value of $Q_0$ is an indication of the area of applicability of a given
parameterisation. Within a common area of applicability, fits with a better
$\chi^2$/dof are to be preferred. If the parameters of a fit have
physical meaning, then their values must be stable when one restricts
the fit to a sub-set of the full data set, or if one limits the area
of applicability, {\it e.g.} but modifying the starting $Q_0$ of a
DIS fit. Similarly, fits that use a handful of parameters are usually
preferred to those that use many. All these features can be studied
numerically, and details can be found in refs.~\cite{PRD2002,JRC}. The use
of these indicators then enables one to decide which model may
be preferred to describe some set of data.

\subsubsection*{Web}
We have also developed an automatic generation of results which are
then gathered in postscript files available on the web
\cite{websummary}, as well as a calculational interface that
predicts values of observables for the first few best parameterisations
\cite{calculator}.
Furthermore, computer-readable files \cite{card}, as well as Fortran code for
the
best models \cite{fortran}, are also given. As we shall see, this is only a
first step
as a full interface between different objects of knowledge still needs
to be built.

\subsection*{Physics}
\subsection*{Soft Forward data: FORWARD2.1}
We started our activities a few years ago \cite{start}, concentrating on
analytic
fits to total cross sections and to the $\rho$ parameter. Such studies
first revealed a few problems with the data set, and then proved the equivalence
of simple, double and triple pole parameterisations in the region
$\sqrt{s}\geq 9$ GeV. This resulted in the first version of the object
of knowledge concerned with soft forward physics, FORWARD1.0.
Its second version \cite{PRD2002}, dating
from last year, came when it was realized that some fits could be extended down
to $\sqrt{s}\geq 4$ GeV. The latter models thus became favored,
and constitute the second version of the object of knowledge, FORWARD2.0.
It now
contains 3092 points (742 above 4 GeV), and 37 adjusted and ranked
models. We have recently used it to produce predictions at present and
future colliders \cite{PRL2002},
and included cosmic ray data to obtain FORWARD2.1,
which is detailed in J.R. Cudell's contribution to these proceedings.

This object of knowledge has demonstrated that $\rho$ parameter data
were poorly reproduced. Some experiments at low energy seem to have
systematic shifts with respect to other experiments, and the $\chi^2/dof$ of
the $\rho$ data is very bad for some data sets ($pp$, $p\pi^+$, $pK^-$).
Although some of the problems can be understood as coming from the
use of derivative dispersion relations (see O.V. Selyugin's contribution to
these proceedings),
discrepancies between data nevertheless make a good fit impossible.

The
only clean way out is to perform the experimental analysis again,
or part of it, either through a check (and correction) of the
theoretical input used, or through a re-analysis of the data in
the Coulomb-nuclear interference region. One thus needs a common
parameterisation of
electromagnetic form factors, a common procedure to analyse data in
the Coulomb-nuclear interference region, a common set of strong
interaction elastic scattering
parameterisations, and a common study of Regge trajectories.
The next few objects of knowledge are devoted to
the systematization of such information.

\subsection*{Regge trajectories: RT1.0$\beta$}
First of all the long way of modelling the forms of Regge trajectories 
for positive and negative values of $t$ should be systemized. Even the 
good old idea of linearity at positive $t$ should be tested and 
maintained.

\begin{figure}[h]
\centerline{\epsfxsize=135mm \epsfbox[14 14 731 525]{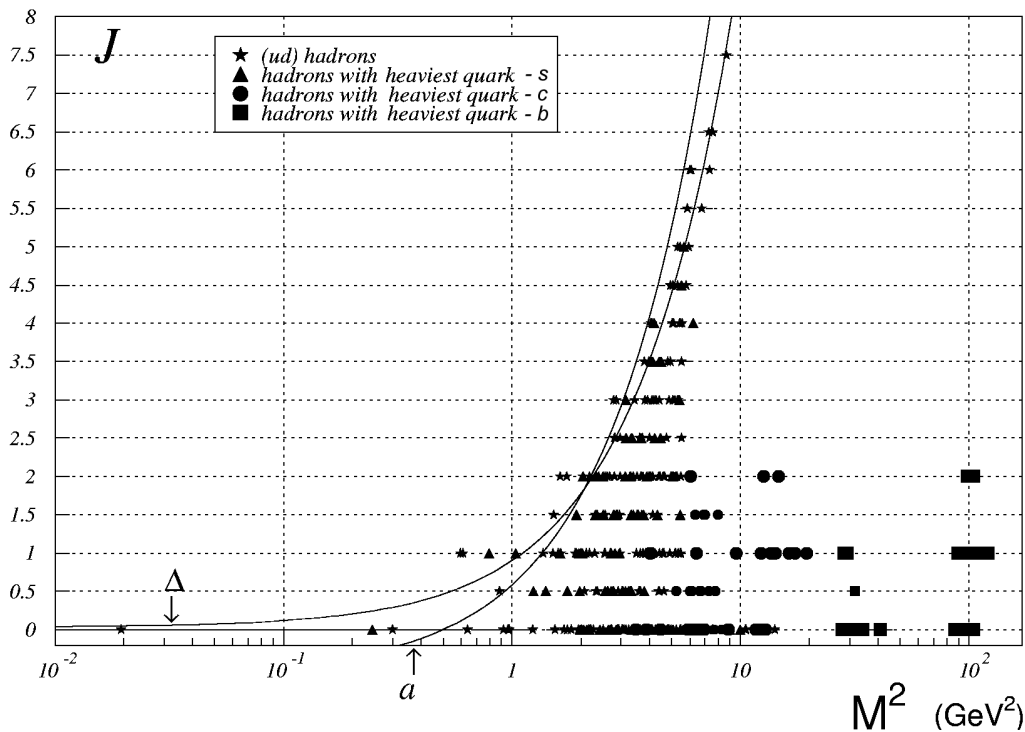}} 
\vspace*{-3mm} 
\caption {Chew-Frautschi plot for all hadrons from RPP-2002}
\label{alls}
\end{figure}

We have extracted from the RPP-2002 database a new set of hadronic 
states (213 mesons and 123
baryons), including their masses, widths and quantum numbers. Corresponding
isotopic multiplets (one isomultiplet -- one point)
are all presented in a log-linear Chew-Frautschi  plot with different 
markers for different flavors to show the similarities and  differences 
of the $(M^2,J)$ populations for different hadron classes, 
see Fig.\ref{alls}. 

 There is
only one linear meson trajectory $(a_2,\quad a_4,\quad a_6)$ that 
give the acceptable fit
quality with weights constructed from the errors in masses. This trajectory
is placed on the Fig. \ref{alls} together with longest baryon trajectory (5 $\Delta$ members) with good fit quality.
\vspace*{1mm}

Preliminary fits of the RPP-2000 data to linear trajectories (in the approximation
where weights in the fits are constructed from $\Delta(M^2) = M \Gamma$,
instead of $\Delta(M^2) = 2 M \Delta M$ ) show a clear systematic
flavor dependence of the slope for mesons, as shown in Table I.
Such a dependence of the slopes on flavor does not seem to be present
 in the baryon case.
 
\begin{figure}[h]
\centerline {\epsfxsize=135mm \epsfbox[0 0 709 709]{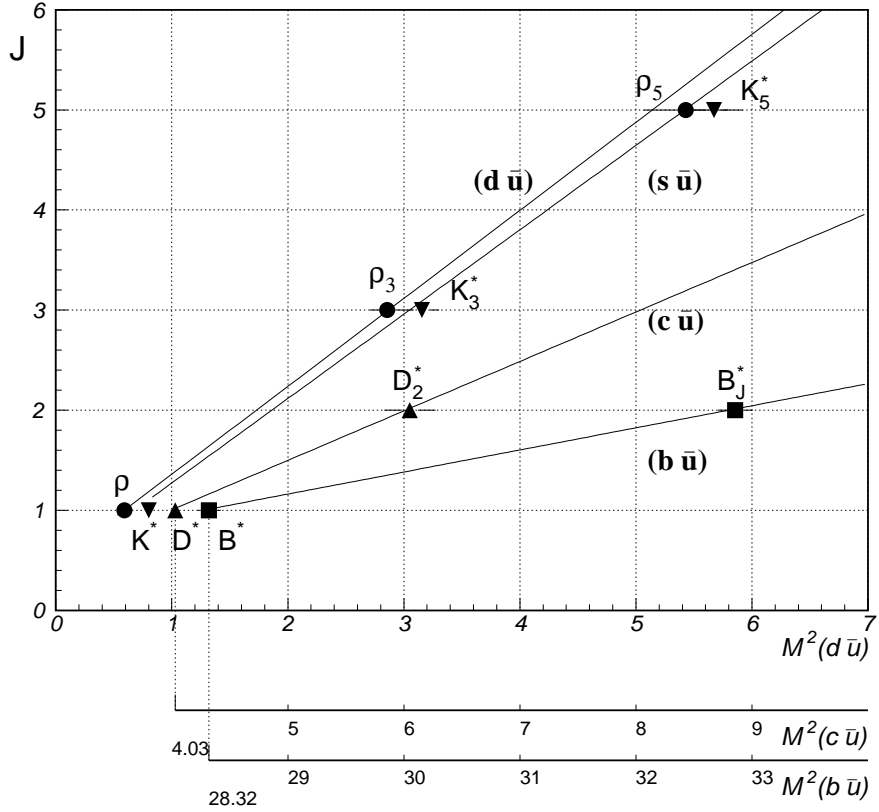}}
\vspace*{-25mm}
\caption{Linear Regge trajectories slopes for different meson flavors}
\label{flavs}
\end{figure}

\vspace*{-8mm}
\begin{center}
\begin{tabular}{ccccc}
\multicolumn{5}{c}{Slopes of meson Regge trajectories as a function of}\\
\multicolumn{5}{c}{their flavor content (obtained from the 2000 data) }\\
\hline
         &q             &s             &c               &b\\
\hline
$\bar q$ &$0.84\pm 0.09$&$0.86\pm 0.02$&$0.49\pm 0.08$  &$0.22\pm 0.01$\\
$\bar s$ &              &$0.82\pm 0.01$&$0.55 \pm 0.01$ &$0.22\pm 0.02$\\
$\bar c$ &              &              &$0.40 \pm 0.01$ &\\
$\bar b$ &              &              &                &$0.11\pm 0.01$\\
\hline
\end{tabular}
\end{center}
We plan to
reiterate fits in this approximation on the 2002 RPP data to see if
the regularity is stable and we will then proceed to collect and 
compare different functional forms of the trajectories on a regular 
basis using the: spectroscopic data together with the elastic 
scattering data; data on two body reactions; decay properties data 
to see if the decision rule for the dichotomy ``quark model hadrons
-- exotic hadrons" can be constructed.

\newpage
\subsection*{Electromagnetic form factors of hadrons: EFFH1.0$\beta$}
This object of knowledge is also under construction. So far only nucleon
emff were considered.
Its data set consists of
785 values of $d\sigma_{ep}/d\Omega$, 29 values of $G_e/G_m$ and 31
values of $\sigma_{tot}$ for ${\overline p} p \to e^+  e ^-$, for a total of 845 points. We ignore derived data
 on emff and produce fits only to the directly measured observables and then
 compare fits. 
   The base of models consists of 4 adjusted and maintained parameterizations.
 
 Recently the extended Gari-Kruempelmann \cite{Gari:1992qw} parameterization
 for the nucleon emff were fitted \cite{Lomon:2001ga} to the most 
 complete data set of the derived data with inclusion of the new 
 data on  $G_E/G_M$ \cite{Jones:1999rz}. To include this extension of the
 Gari-Kruempelmann parameterization to the model base we started to check
 if it could be reasonably fitted to our database.
 
  It turns out that in the VMD part of parameterization it is enough to include only one vector meson ($\rho(770)$) to obtain a reasonable fit to the $d\sigma/dt$ and  $G_E/G_M$ data. 

However it leads to the determination of the electric and magnetic radii
of the nucleons that are incompatible with that determined from the Lamb
shift in hydrogen atom measurements.

\begin{table}[h]
\begin{center}
\caption{Fit to the $d\sigma/dt$ data: $\chi^2/d.o.f. = 0.91$}
\begin{tabular}{crr|rrrr}
\hline
$\langle r^2 \rangle$~$fm^2$ &Value&$\sigma^2$~$fm^2$ &\multicolumn{4}{|c}{Correlations}\\
\hline
$\langle (r_E^p)^2 \rangle$& 0.6906&2.7E-03&  1.00  &$-0.01$&$0.22$&$-0.26$\\
$\langle (r_M^p)^2 \rangle$& 0.6926&3.1E-03&$-0.01$&  1.00  &$-0.44$&$0.95$\\
$\langle (r_E^n)^2 \rangle$&$-$0.4266&3.2E-03&$0.22$&$-0.44$&  1.00  &$-0.65$\\
$\langle (r_M^n)^2 \rangle$& 0.9003&5.3E-03&$-0.26$&$0.95$&$-0.65$& 1.00\\
\hline 
\end{tabular}
\end{center}
\label{TabII}
\end{table}

\vspace*{-5mm}

\begin{table}[h]
\begin{center}
\caption{Fit to the $d\sigma/dt$ and $G_E/G_M$ data: $\chi^2/d.o.f. = 1.03$}
\begin{tabular}{crr|rrrr}
\hline
$\langle r^2 \rangle$~${\rm fm^2}$&Value&$\sigma^2$~${\rm fm^2}$&
\multicolumn{4}{|c}{Correlations}\\
\hline
$\langle (r_E^p)^2 \rangle$& 0.6650  &1.7E-03&  1.00  &$ 0.32 $&$-0.24$&$0.69$\\
$\langle (r_M^p)^2 \rangle$& 0.7153  &4.7E-03&$0.32$&  1.00  &$0.73$&$0.28$\\
$\langle (r_E^n)^2 \rangle$&$-$0.3411&15.6E-03&$-0.24$&$0.73$&  1.00  &$-0.46$\\
$\langle (r_M^n)^2 \rangle$& 0.8965  &4.7E-03&$0.69$&$0.28$&$-0.46$&1.00\\
\hline 
\end{tabular}
\end{center}
\label{TabIII}
\end{table}

\noindent
We see from  Tables \ref{TabII}, \ref{TabIII} that estimates of
the physical parameters changed
markedly with the addition of new observables measured in the same range of kinematic
variables.

It should be noted that the mean square proton radii
$\langle (r_E^p)^2 \rangle = 0.61$~${\rm fm^2}$, 
$\langle (r_M^p)^2 \rangle = 1.82$~${\rm fm^2}$
 calculated from the fit
obtained in \cite{Lomon:2001ga} for the same model but with a VMD part containing
$\rho$, $\omega$, and $\phi$ contributions are even worse in comparison with
estimates from the Lamb shift data.  This is a signal for possible problems with the database and/or with parameterizations (see also \cite{Brash:2001qq}).  Further cross-assessment iterations with models ranking are needed. 
 
\subsection*{Forward elastic scattering of hadrons: FESH1.0$\beta$}
This database contains the measured differential distribution $d\sigma/dt$ for
$|t|<0.6$ GeV$^2$
for $\pi^\pm p$ (438 points at 73 energies), $K^\pm p$ (204 points at
34 energies) and $\bar p p$, $pp$ (564 points at 94 energies). The associated
object of knowledge is interfaced with the FORWARD, EFFH and RT
objects of knowledge as
\begin{equation}
{d\sigma\over dt} = \pi \left(|f_c|^2+2Re(f_c^* f_h)+|f_h^2|^2\right)
\end{equation}
with $f_c$ the Coulomb amplitude, which depends on the form factors
of EFFH, $f_h$ the hard-interaction amplitude, which depends on
$\sigma_{tot}$ and $\rho$ (from FORWARD), and on Regge trajectories (from RT).

\begin{figure}[h]
\vspace*{-5mm}
\centerline {\epsfxsize=135mm \epsfbox[28 28 749 929]{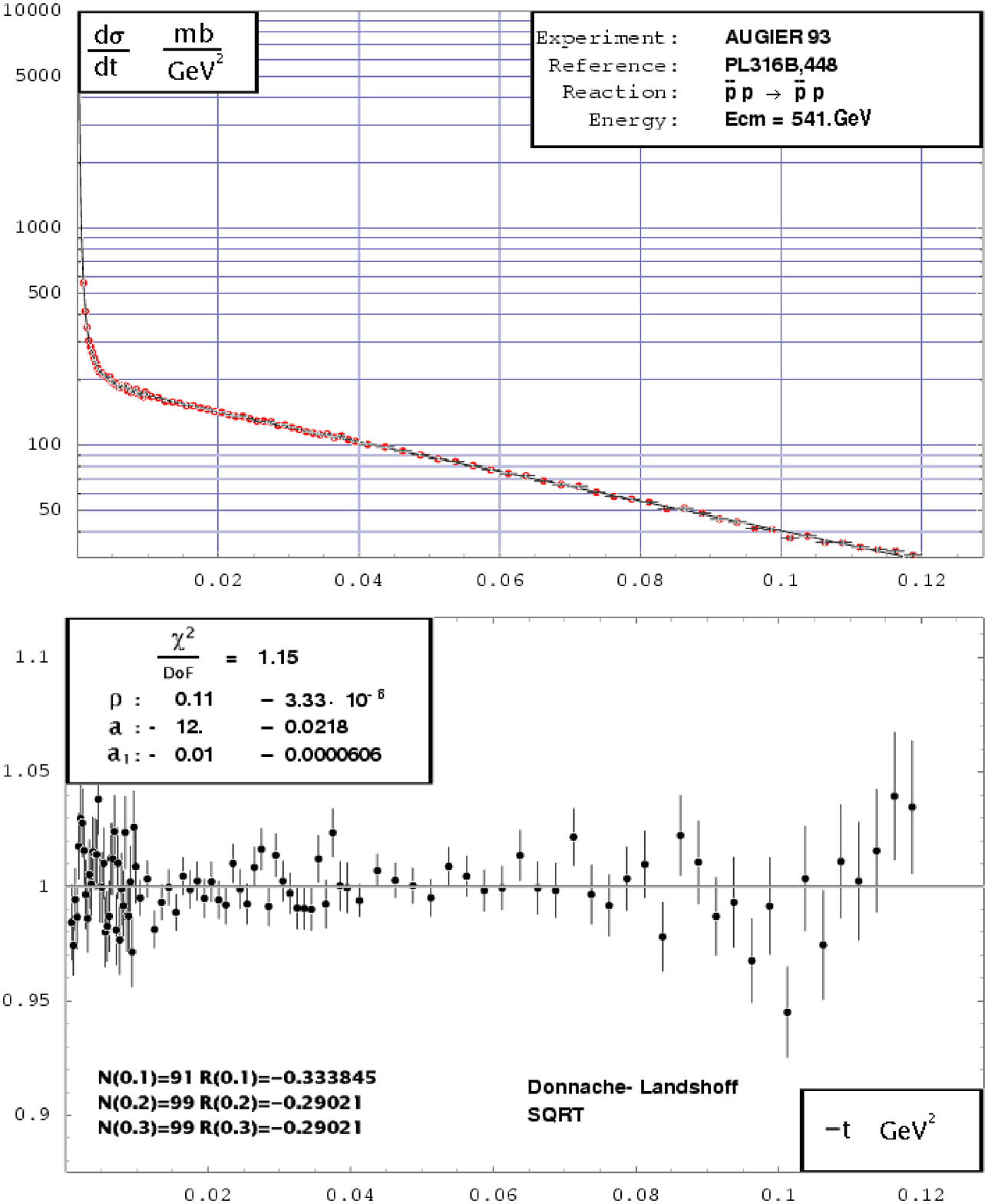}} 
\vspace*{-5mm}
\caption{Fit to the $d\sigma/dt$ and results of autocorrelation function calculation}
\label{agierbaev}
\end{figure}

So far, we have done a preliminary study trying to find regularity in 
energy of the several claimed  evidences for  
oscillations on the diffraction cone. The method to reveal the oscillations
is illustrated in the Figure \ref{agierbaev}
  
 Using the na\"\i ve models for the diffractive
cone description $( A(s)e^{B(s)\alpha(t)})$ and the standard Coulomb amplitude with popular dipole(pole)
charge form-factors for nucleons(mesons), we calculate the normalized autocorrelation $R(s)$ of the difference
$T(s,t) = ({{d\sigma^{data}/dt} \over {d\sigma^{theory}/dt}} -1)$ for  195 experimental distributions $d\sigma/dt$ at different values of $P_{lab} \ge 10$ GeV/$c$ and having more
than 7 data points in the region $|t|<0.6$
GeV$^2$.

$$ R(s) = {\sum_i {{T(s,t_i)T(s,t_{i+1})} \over {\sigma_i^2\sigma_{i+1}^2}}}
\left ( {\sum_i {1 \over {\sigma_i^2\sigma_{i+1}^2}}}\right )^{-1},$$

\noindent
where $\sigma_i = {\sigma^{data}_i \over {d\sigma^{theory}(s,t_i)/dt}}$.

 Large $( > 1)$ values of the autocorrelator are signals for oscillations
or fit biases, large negative values $( < -1)$ are signals that we have
problems with data (the errors are over-estimated).
 
\begin{figure}[h]
\centerline {\epsfxsize=135mm \epsfbox[0 0 612 792]{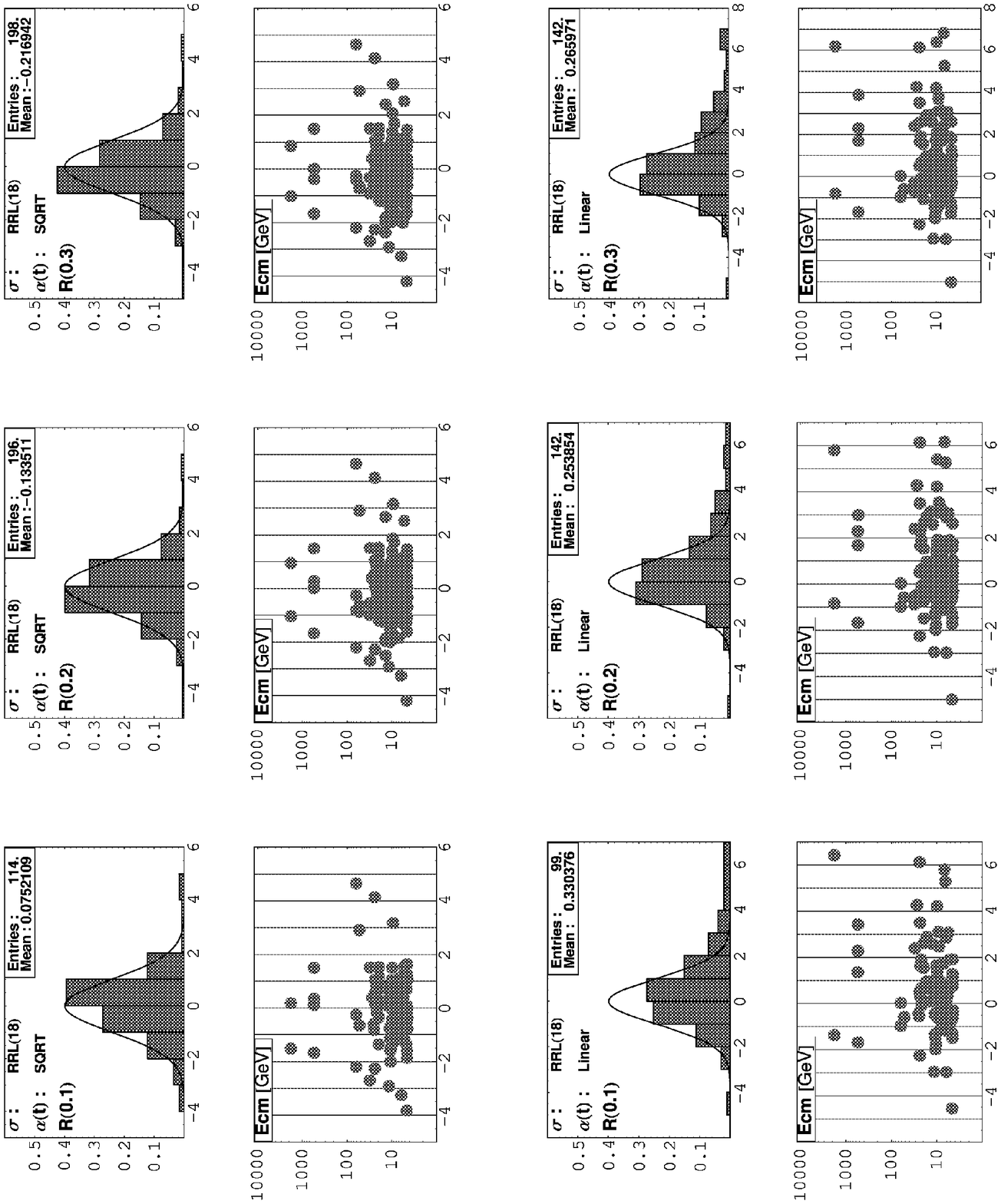}}
\vspace*{-10mm}
\caption{``Statistical pattern" of the autocorrelations as the indicator
for the fine structure on the diffractive cone}
\label{statview}
\end{figure}

 Figure \ref{statview} shows that almost all values of the 
autocorrelations for the three intervals $|t| < 0.1$ GeV$^2$, $|t| <0.2$
GeV$^2$, and $|t| < 0.3$ GeV$^2$  are close to the normal distribution. There are some outstanding
points but the data on the corresponding scatter plots do not show 
any stable regularity in energy dependence of the scatter plots structures.
These outstanding autocorrelation values may be due hidden t-dependent 
systematic effects, or to biases in the fits on the diffraction cone.
 
Furthermore, the possible oscillation pattern seems to be 
model-dependent, as seen on the same figure.
 For example, the Figure \ref{agierbaev} clearly show absence of
the oscillations claimed in some phenomenological papers. Our preliminary
conclusion is based on the analyses of two different forms of the t-dependence
of the pomeron trajectory: linear and square root dependence with branch
point at $t=4m_{\pi}^2$. The optical points also
were calculated with use different models of the energy dependence of the
total cross sections.

 To make unambiguous conclusions we need more iterations
of cross-assessments to find a parameterization that give good description
of the diffractive cones and their evolution with energy. Having such a 
parameterisation it will be possible to clarify the situation with claimed
oscillations.   

\subsection*{Cross section in $e^+e^- \to hadrons$, $R$, QCD tests: CSEE1.0}
As a parallel activity, we have gathered \cite{hepph_ee} the data for the
annihilation
cross sections $\sigma_{e^+e^- \to hadrons}$ and for their ratio $R$
to $\sigma_{ee \to \mu\mu}$ for $0.36$ GeV $ \le \sqrt{s} \le 188.7$ GeV. The database consists of 1066 points rescaled to the hadronic
$R$. The QCD fit to the hadronic part clearly shows that a 3-loop
calculation is preferred with respect to the na\"\i ve Born formula, 
and leads to the
following value of $\alpha_S$:
\begin{equation}
\alpha_S(M_Z^2)=0.128\pm 0.032
\end{equation}

\begin{figure}[h]
\centerline{\hspace*{3mm}\epsfxsize=120mm \epsfbox[40 0 607 283]{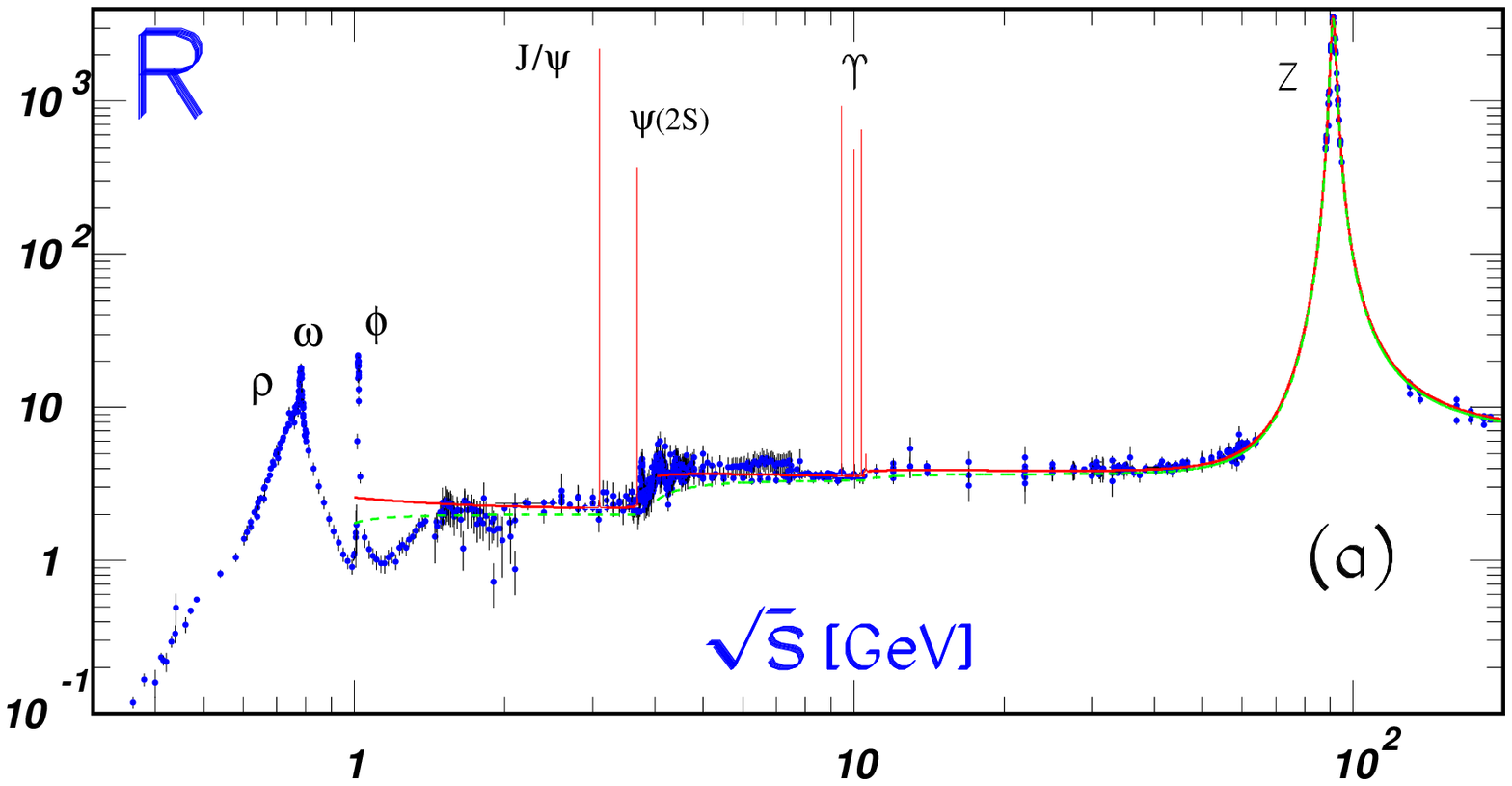}} \centerline{\epsfxsize=115mm \epsfbox[50 -10 900 387]{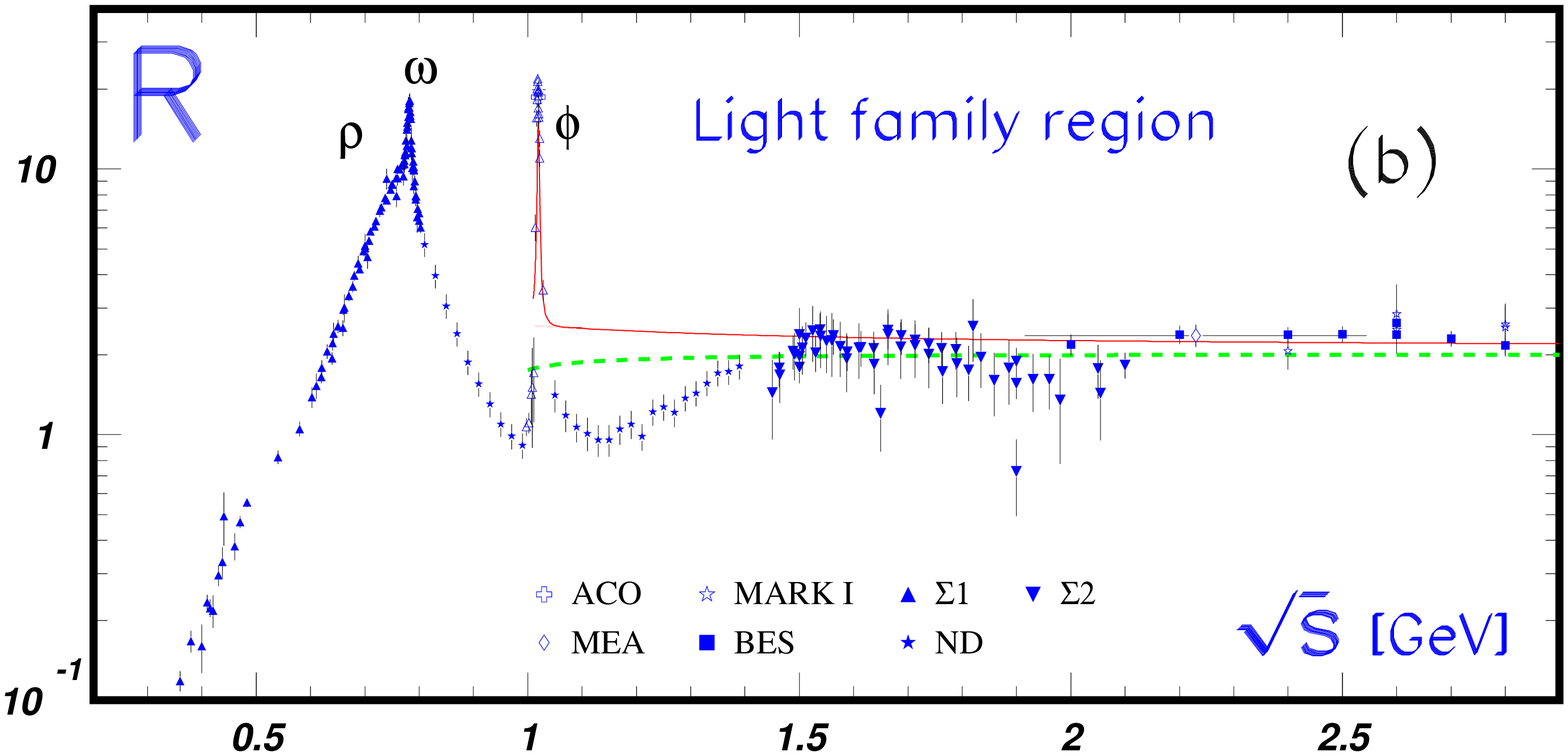}}
\centerline{\epsfxsize=115mm \epsfbox[40 -10 890 273]{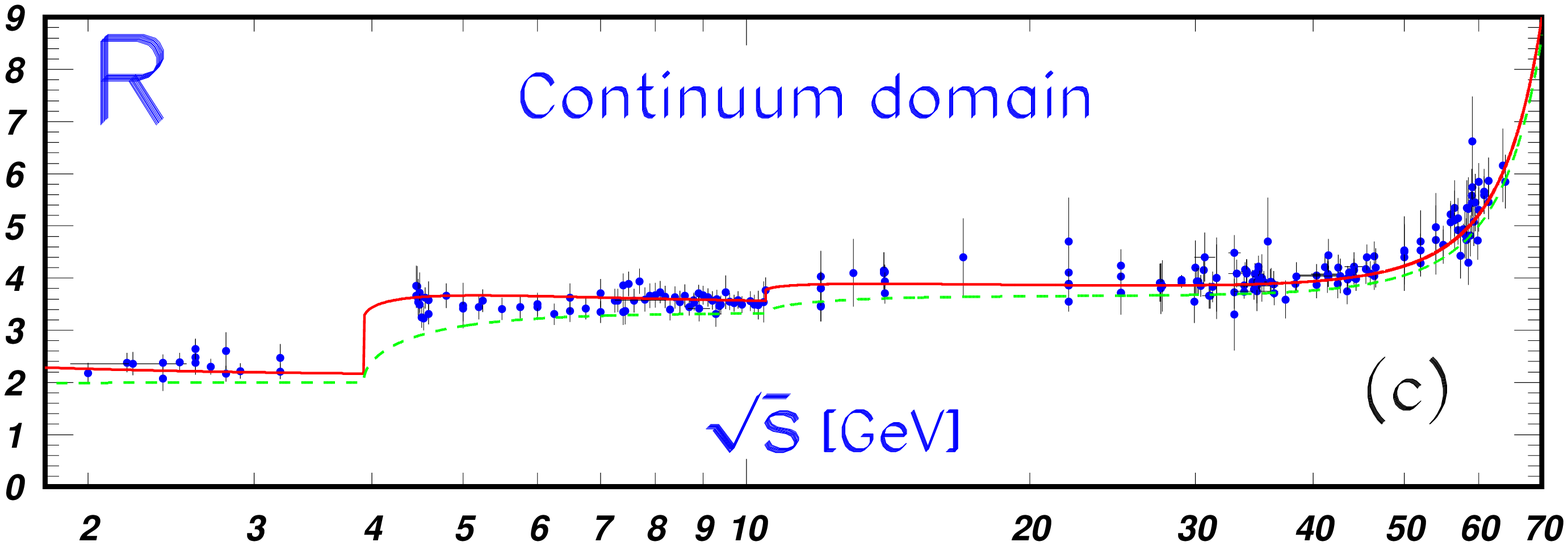}}
\caption{ {\bf\rm (a)} World data on the ratio 
$R = \frac{\sigma(e^+e^- \to q\bar q \to {\mathrm hadrons})}
{\sigma(e^+e^- \to \mu^+\mu^-)}$. 
{\bf\rm (b)} Low $\sqrt{s}$ region crucial for the
evaluation of 
$\Delta\alpha^{had}_{QED}(M_Z)$, ${a_\mu}^{had}_{LO}$, {\it etc}.
{\bf\rm (c)} Applicability  domain of 3-loop pQCD.
Solid curves are 3-loop pQCD predictions (plus Breit-Wig\-ner
for narrow resonances on {\rm (a)} and {\rm (b)}). 
Broken curves show the ``na\"\i ve'' parton model prediction.
Masses of $c$ and $b$ quarks are taken into account.
The full set of radiative corrections is applied to all data.
Further details and references to the original experimental data 
can be found in \cite{hepph_ee}.
See also \cite{RPP2002_ee}.
}
\label{Fig_ee}
\end{figure}

All available data on the total cross section and the 
$R$ ratio of $e^+e^- \to {\mathrm hadrons}$
are compiled from the {\rm PPDS(DataGuide, ReacData)}
(IHEP, Prot\-vino, Russia) 
and {\rm HEPDATA(Reaction)} (Durham, UK) 
databases and transformed to a compilation of data on the ratio 
$R = \frac{\sigma(e^+e^- \to q\bar q \to {\rm hadrons})}
{\sigma(e^+e^- \to \mu^+\mu^-)}$, with the full set of radiative corrections.
This compilation is the most complete set of evaluated hadronic $R$ ratio data 
publicly available to date.
The current status of the data is shown in Fig.~\ref{Fig_ee}.
The compilation is continuously maintained so that
new experimental data are added as they become
available.

The compilation is intended for tests of pQCD calculations 
as well as for a precise evaluation of  
hadronic contributions to $\Delta\alpha_{QED}(M_Z)$, $a_\mu=(g_\mu-2)$, 
{\it etc}. 
The results we obtained so far  are as follows:
current theoretical predictions from the parton model and pQCD
are well supported by the world ``continuum'' data on 
$\sigma_{tot}(e^+e^- \to {\mathrm hadrons})$.
Our preliminary value of $\Delta\alpha^{had}_{QED}(M_Z)$ 
is $0.02736 \pm 0.00040{\mathrm (exp)}$,
in agreement with the results of other groups \cite{RPP2002}. 
The refinement of the $\Delta\alpha^{had}_{QED}(M_Z)$ calculation
and the evaluation of ${a_\mu}^{had}_{LO}$ are in progress.

Computer-readable data files are accessible on the Web at

{\rm\tt http://pdg.lbl.gov/2002/contents\underline{~}plots.html} 
(see also \cite{RPP2002_ee}) and

{\rm\tt http://wwwppds.ihep.su:8001/eehadron.html}

\section*{Prospects for the future}
The various objects of knowledge described in this report should be
released to the community within a year. The FORWARD object of knowledge  will
also probably be renewed in light of the analysis of $\rho$. We also
plan soon to build objects of knowledge devoted to 2-body processes
at large $s$ and $t$, to photoproduction of vector mesons, and to
hadronic multiplicities.

\section*{Acknowledgments }
The COMPAS group was supported in part by the Russian Foundation for
Basic Research grants RFBR-98-07-90381 and RFBR-01-07-90392. K.K.
is in part supported by the U.S. D.o.E. Contract DE-FG-02-91ER40688-Task~A. E. Martynov and O. Selyugin are visiting fellows of the Belgian FNRS.

\end{document}